# $\theta$ Vacua, Confinement and the Continuum Limit [*]

G. Schierholz

Deutsches Elektronen-Synchrotron DESY, D–22603 Hamburg, Germany
and
Gruppe Theorie der Elementarteilchen, Höchstleistungsrechenzentrum HLRZ,
c/o Forschungszentrum Jülich, D–52425 Jülich, Germany

We investigate the dependence of the $CP^3$ model and of the $SU(2)$ Yang-Mills theory on the vacuum angle $\theta$. The $CP^3$ model exhibits a first order deconfining phase transition in $\theta$. The critical value of $\theta$ runs from $\pi$ in the strong coupling limit towards zero as $\beta$ is taken to infinity. Qualitatively the same behavior is found in the $SU(2)$ Yang-Mills theory in four dimensions. We discuss the renormalization group trajectories. Only the line with $\theta = 0$ allows the cut-off to go to infinity, while the lines with $\theta > 0$ end on the line of first order phase transitions. Thus $\theta$ is forced to zero in the continuum limit.

## 1. INTRODUCTION

In QCD, and theories of its kind, the proper vacuum states are superpositions of vacua of different winding numbers $n$:

$$|\theta\rangle = \sum_n \exp(i\theta n)|n\rangle. \qquad (1)$$

These so-called $\theta$ vacua are realized by adding a CP violating term to the action,

$$S_\theta = S - i\theta Q, \qquad (2)$$

where $S$ is the standard action and $Q$ is the topological charge. A priori $\theta$ is a free parameter. Since no CP violation has been observed in the strong interactions, $\theta$ must however be very close to zero. This is known as the strong CP problem.

The QCD vacuum can be understood as a dual superconductor [1] in which color magnetic monopoles condense and color electric charges, i.e. quarks and gluons, are confined by a dual Meissner effect. This picture has been successfully tested in lattice simulations [2]. In the $\theta$ vacuum the monopoles acquire a color electric charge of the magnitude $\theta/2\pi$ [3]. For $\theta \neq 0$ one would then expect that the long-range color electric forces are screened by monopoles and that confinement is lost. This suggests that QCD is a viable theory only for $\theta = 0$.

What makes a lattice investigation at non-vanishing values of $\theta$ very difficult is the fact that the action is complex. This has led us to study the $\theta$-dependence in a simple model first.

## 2. THE $CP^3$ MODEL

A class of models, which in many respects are similar to QCD, are the $CP^{N-1}$ models in two space-time dimensions. They describe N-component, complex scalar fields $z_a(x)$ of unit length which interact minimally with the composite gauge field $A_\mu(x) = \frac{i}{2}\bar{z}_a(x)\overleftrightarrow{\partial}_\mu z_a(x)$. The topological charge is constructed from the gauge field in the usual way.

The $\theta$-dependence of the theory is governed by the partition function

$$Z(\theta) = \sum_Q \exp(i\theta Q)\,p(Q) \equiv \exp(-VF(\theta)), \qquad (3)$$

where $p(Q)$ is the probability of finding a field configuration with charge $Q$ and $F(\theta)$ is the free energy per space-time volume $V$. In terms of $F(\theta)$ the average topological charge density is given by

$$\frac{1}{V}\langle\theta|Q|\theta\rangle \equiv -iq(\theta) = -i\frac{dF(\theta)}{d\theta}, \qquad (4)$$

and the string tension of two external particles of charge $e$ and $-e$ (in units of the intrinsic fundamental charge) turns out to be

$$\sigma(e,\theta) = F(\theta + 2\pi e) - F(\theta). \qquad (5)$$

We have chosen to examine the $CP^3$ model [4] which is not expected to be affected by dislocations.

---

[*] Talk given at *International Symposium on Lattice Field Theory*, Bielefeld, 1994.

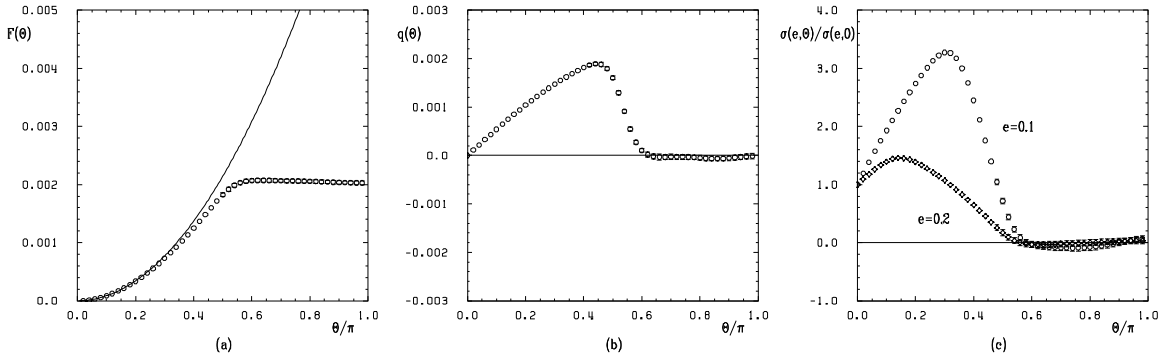

Figure 1. The free energy $F(\theta)$ (a), the charge density $q(\theta)$ (b) and the string tension $\sigma(e,\theta)$ (c) as a function of $\theta$ for the $CP^3$ model on the $V = 64^2$ lattice at $\beta = 2.7$. The solid curve in (a) is the prediction of the large-N expansion. Only the first half of the $\theta$ interval is displayed.

We expect to find a first order phase transition in $\theta$ from a confining phase to a Higgs phase. This will manifest itself in a kink in $F(\theta)$ and a discontinuity of its first derivative.

In Fig. 1 I show $F(\theta)$ on the $V = 64^2$ lattice at $\beta = 2.7$. We see a kink at $\theta = \theta_c \approx 0.5\,\pi$. I also show the topological charge density on the same lattice. This can be interpreted as a background electric field. We see that $q(\theta)$ increases linearly with $\theta$ up to $\theta = \theta_c$, where it jumps to zero, indicating that the phase transition is accompanied by a collapse of the background electric field due to pair creation. To demonstrate that the phase transition is a deconfining phase transition I show the string tension in Fig. 1 as well, again on the same lattice, for two different external charges. We see that the string tension vanishes for $\theta \geq \theta_c$. This is a consequence of the fact that $F(\theta)$ is constant in that region.

Only on an infinite lattice and at $\beta = \infty$ can we expect that $\theta_c = 0$. We have done calculations at $\beta = 2.5, 2.7$ and $2.9$ on lattices ranging from $28^2$ to $200^2$. For a first order phase transition we expect $\theta_c(\beta, V) - \theta_c(\beta, \infty) \propto V^{-1}$. We find our results to be in good agreement with this behavior. This allows us to extrapolate the lattice data to the infinite volume. The final result is that

$$\theta_c(\beta, \infty) \propto \xi^{-1} \to 0 \qquad (6)$$

as $\beta \to \infty$, where $\xi$ is the correlation length. ($\xi \approx 20$ at our largest $\beta$ value.) This is to be contrasted with the strong coupling result $\theta_c = \pi$ [5].

We have also investigated the $\beta$-function. We find that it does not depend on $\theta$ in the region $\theta < \theta_c$, where it agrees very well with the two-loop perturbative result, while at the phase transition it appears to drop to zero.

## 3. THE SU(2) YANG-MILLS THEORY

As the next step we have investigated the SU(2) Yang-Mills theory in four dimensions. We chose the action

$$S = \sum_p [\beta_f (1 - \frac{1}{2} \text{Tr} U_p) + \beta_a (1 - \frac{1}{4} |\text{Tr} U_p|^2)], \quad (7)$$

and we used Phillips and Stone's algorithm for the topological charge [6]. This combination evades problems with dislocations for $\beta_a/\beta_f < -0.3$ [7]. So far we have done calculations [8] at $\beta_f = 3.2$, $\beta_a/\beta_f = -0.32$ on the $12^4$ lattice. (This corresponds to $\beta \approx 2.4$ for the Wilson action.) In Fig. 2 I show $F(\theta)$. We see a kink at $\theta = \theta_c \approx 0.2\,\pi$, which indicates a first order phase transition. At strong coupling we obtain analytically $\theta_c = \pi$. We may therefore expect to find the same picture as in the $CP^3$ model.

## 4. THE PHASE DIAGRAM

Our results suggest the phase diagram shown in Fig. 3. There exists a line of first order phase transitions, separating the confining phase from a deconfining phase, which starts at $\theta_c = \pi$ in the strong coupling region and then falls off rapidly towards zero as $\beta$ is taken to infinity. At intermediate values of $\beta$ this line may not be universal and depend on the particular form of the action and the charge algorithm. But we expect the results to be unique for larger values of $\beta$.

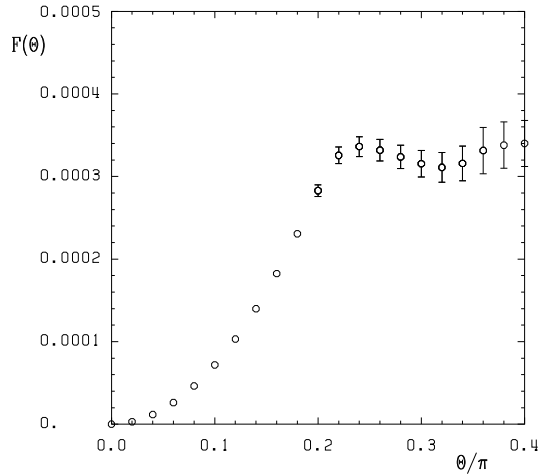

Figure 2. The free energy $F(\theta)$ as a function of $\theta$ for the SU(2) Yang-Mills theory on the $12^4$ lattice at $\beta_f = 3.2$, $\beta_a/\beta_f = -0.32$.

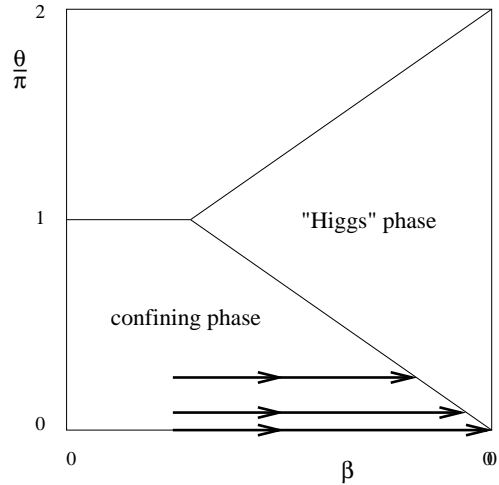

Figure 3. The phase diagram. The horizontal lines marked by arrows are the lines of constant physics.

What does this imply for the continuum limit? As already mentioned, we found that the $\beta$-function is independent of $\theta$ for $\theta < \theta_c$. The CP-violating observables can be written as $\theta$ times an observable not depending explicitly on $\theta$, plus similar higher order expressions. $\theta$ itself is a renormalized quantity. We therefore expect that the lines of constant physics run parallel to the axis $\theta = 0$, as is indicated by the lines marked by arrows in Fig. 3. For $\theta > 0$ these lines end on the line of first order phase transitions, which results in a finite cut-off. Only the line $\theta = 0$ will run into the (tri-)critical point $(\beta,\theta) = (\infty,0)$, where the cut-off can be taken to infinity. Thus, as the correlation length goes to infinity, $\theta$ goes to zero.